\documentclass[aps,pra,epsfigure,twocolumn,showpacs]{revtex4}
\usepackage{amsmath}
\usepackage{amstext}
\usepackage{latexsym}
\usepackage{psfig}
\usepackage{graphicx}
\usepackage{amsfonts}

\newcommand{\ket}[1]{\left\vert#1\right\rangle}
\newcommand{\modul}[1]{\left\vert#1\right\vert}

\newcommand{\one}{\mbox{$1 \hspace{-1.0mm}  {\bf l}$}}
\newcommand{\gate}[1]{{\textsf {#1}}}
\newcommand{\pro}[2]{\left\vert#1\rangle\langle#2\right\vert}
\newcommand{\cnot} {{\textsf {CNOT}}}

\newcommand{\bra}[1]{\left\langle#1\right\vert}
\newcommand{\sand}[3]{\left\langle#1\vert#2\vert#3\right\rangle}

\begin{document}

\title{Non-Local Quantum Gates: a Cavity-Quantum-Electro-Dynamics implementation}
\author{M. Paternostro}
\affiliation{School of Mathematics and Physics, The Queen's University,
Belfast BT7 1NN, United Kingdom}
\author{M. S. Kim}
\affiliation{School of Mathematics and Physics, The Queen's University,
Belfast BT7 1NN, United Kingdom}
\author{G. M. Palma}
\affiliation{Dipartimento di Tecnologie dell'Informazione,
Universita' di Milano,
Via Bramante 65,
26013 Crema, Italy}
\affiliation{}
\affiliation{NEST \& INFM}
\date{\today}

\begin{abstract}

The problems related to the management of large quantum registers could be handled in the context of distributed
quantum computation: unitary non-local transformations among spatially separated local processors are realized
performing local unitary transformations and exchanging classical communication. In this paper, we propose a
scheme for the implementation of universal non-local quantum gates such as a controlled-$\gate{NOT}$
($\cnot$) and a controlled-quantum phase gate ($\gate{CQPG}$). The system we have chosen for their physical
implementation is a Cavity-Quantum-Electro-Dynamics (CQED) system formed by two spatially separated microwave
cavities and two trapped Rydberg atoms. We describe the procedures to follow for the realization of each step
necessary to perform a specific non-local operation.
\end{abstract}
\maketitle


\section{Introduction}

\begin{figure}[b]
\centerline{\psfig{figure=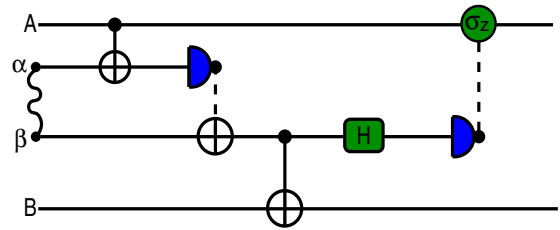,width=7.3cm,height=3.0cm}}
\caption{Quantum circuit for a non-local $\cnot$ gate realized using a
shared ebit and two classical bits of communication. In this scheme, $A$ is
the control qubit and $B$ is the
target qubit. The joined state of qubits $\alpha$ and $\beta$ encodes the needed ebit
(entanglement is represented by the wavy line). Classical communication is
represented by dashed lines and symbols for $\sf CNOT$ operation, Hadamard transform and Pauli 
$\sigma_z$ operator are shown. The measurements performed on the atomic qubit are schematically represented by detectors.}
\label{Figure1}
\end{figure}

One of the major problems in the experimental implementation of large scale quantum computing devices is
scalability, i.e. the physical control at microscopic level of a large number of quantum subsystems. In
particular the destructive effects of decoherence grow with the size of the register \cite{Suominen}. Furthermore,
undesired interactions among qubits of the same quantum register settle in an uncontrollable
way~\cite{DivLoss,Div}. One  possible solution to this problem could be distributed quantum computing. In this
architecture  a quantum computer is thought as a network of spatially separated devices, which we call {\em
local processors}, each operating on a small number of qubits~\cite{macchiavello}. Such a design of a quantum
computer has recently become particularly stimulating in view of the papers by Eisert {\it et al}.~\cite{Plenio} and
by Collins {\it et al}.~\cite{Popescu}. In these works, the minimal amount of classical and quantum resources needed
to realize a general non-local unitary transformation is investigated. In the case of two-qubits gates, two bits
of classical communication and the maximally  entangled state of a shared pair of qubit ({\it{ebit}}) are proved
to be necessary and sufficient resources to implement a $controlled-U$ gate \cite{Plenio}. In particular, in
ref.~\cite{Plenio}, a theoretical protocol for the optimal implementation of a non-local controlled-$\gate{NOT}$
gate ($\cnot$) is described. This result is relevant since $\cnot$ and single qubit operations constitute an
adequate set for quantum computation~\cite{Preskill,Nielsen,Massimo,Ekert,Vedral}. We have summarized the
protocol, using quantum circuit notation, in Fig.~\ref{Figure1}.

In this paper we propose an experimental scheme for the physical
implementation of a non-local $\cnot$ (according to the protocol proposed in ref.~\cite{Plenio}) and of a
non-local controlled-quantum phase gate ($\gate{CQPG}$) in a Cavity-Quantum-Electro-Dynamics (CQED) set-up.

The paper is structured as follows: in section \ref{protocolli} we describe the protocol of ref.~\cite{Plenio}
for the local implementation of a non-local $\cnot$ and we show how to modify it to obtain a non-local
control quantum phase gate. This latter is a relevant result as well, because the set of quantum gates that comprehends 
$\gate{CQPG}$ and single qubit rotations is universal for quantum computation~\cite{Preskill,Nielsen,Massimo}.
Section \ref{system} is devoted to a brief description of the experimental set-up we
propose in order to implement these non-local gates. A short summary of the features of the interactions of a
two level atom with a single mode of a cavity field is presented. We also describe an approach to the
interaction of the atom with an external classical pulse. In sections \ref{cnotnonlocale} and \ref{qpgnonlocale}
we analyze in detail each step that compose the experimental scheme for the physical realization of the
non-local $\cnot$ and $\gate{CQPG}$. We show how they are realizable in a CQED system including the description
of a procedure for the preparation of the computational register and we comment on the way to obtain the required ebit. 
Every {\em local operation} is analyzed in full detail.


\section{The theoretical protocols}\label{protocolli}

In this section we briefly outline the protocol
proposed by Eisert {\it et al}. \cite{Plenio}. 
We also show how the protocol can be modified to get a non-local $\gate{CQPG}$.

Qubits $A$ and $B$ are, respectively, the control and the target of the non-local $\cnot$
while $\alpha$ and $\beta$ are two auxiliary qubits encoding an ancillary ebit. Alice (Bob) has
access only to qubits $A$ and $\alpha$ ($B$ and $\beta$). We assume that the
initial state of qubits $A$ and $B$ is
\begin{equation}
\ket{\varphi_{in}}_A\otimes\ket{\varphi_{in}}_B=(a\ket{1}+b\ket{0})_A\otimes(c\ket{1}+d\ket{0})
_B
\end{equation}
while the ebit is set in the Bell state
$\frac{1}{\sqrt2}(\ket{01}+\ket{10})_{\alpha\beta}$ (to optimize the
fidelity of the non-local $\cnot$ the
joint state of $\alpha$ and $\beta$ must be a maximally entangled state
\cite{Kimprob}). The protocol can be
read easily from Fig.~\ref{Figure1}. First of all Alice performs a local
$\cnot_{A\alpha}$ where $A$ is the
control and $\alpha$ is the target followed by an orthogonal measurement on
$\alpha$. This transfers
entanglement from qubits $\alpha+\beta$ to qubits $A+\beta$. Bob then uses
the classical information on the measurement result of qubit $\alpha$ to act on qubit $\beta$: if the
outcome of Alice's  measurement is
$\ket{1}_{\alpha}$, Bob applies a $\sf NOT$ on qubit $\beta$ while he
applies $\one$ if Alice detects
$\ket{0}_{\alpha}$. This gives the following state:
\begin{equation}
\begin{split}
&ac\ket{111}_{A\beta{B}}+ad\ket{110}_{A\beta{B}}+\\
&bc\ket{001}_{A\beta{B}}+bd\ket{000}_{A\beta{B}}.
\end{split}
\end{equation}

Now, Bob first applies a $\cnot_{\beta{B}}$ followed by a Hadamard transform on qubit $\beta$ and then detects
its state. Depending on the measurement outcome, the state of $A+B$ is projected onto two different states. If
Bob detects $\ket{0}_{\beta}$, the state of system $A+B$ is exactly what we expect from a $\cnot_{AB}$ applied
on the initial state of the $A+B$ system. If, on the other hand, $\ket{1}_{\beta}$ is detected Alice applies the
Pauli $\sigma_{z}$ operator to qubit $A$, shifting the phase of $\ket{1}_{A}$ by $\pi$ and leaving $\ket{0}_{A}$
unaltered. This gives the desired output state
\begin{equation}
ad\ket{11}_{AB}+ac\ket{10}_{AB}+bc\ket{01}_{AB}+bd\ket{00}_{AB}.
\end{equation}

It is worth pointing out that the protocol works not only for the product
input states of $A+B$ considered here but also for entangled input ones.

With a minor change, the protocol can be modified to implement a non-local
$\gate{CQPG}$ defined in the computational basis of qubits $A$, $B$ as

\begin{equation}
\label{qpgset}
\begin{split}
&\ket{00}\rightarrow\ket{00}\hskip0.5cm\ket{01}\rightarrow{e}^{i\phi}\ket{01}\\
&\ket{10}\rightarrow\ket{10}\hskip0.5cm\ket{11}\rightarrow\ket{11}.
\end{split}
\end{equation}

Indeed it is enough to substitute the $\cnot_{\beta{B}}$ gate shown in
Fig.~\ref{Figure1} with a $\gate{CQPG}_{\beta{B}}$. A straightforward calculation
shows that the overall effect of such
modified circuit is the desired $\gate{CQPG}_{AB}$.


\section{The physical system}
\label{system}

\begin{figure}[b]
\centerline{\psfig{figure=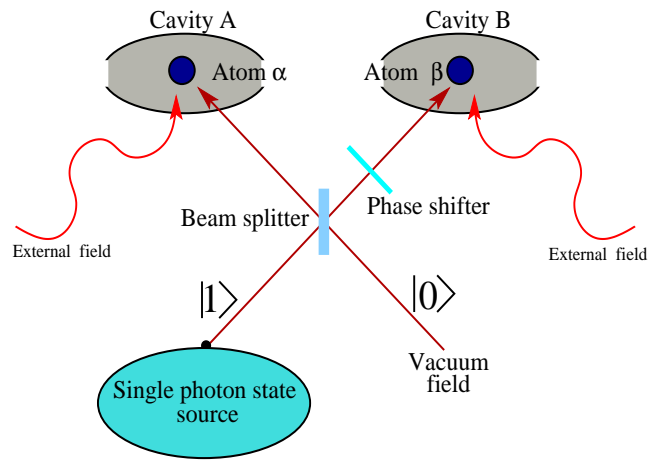,width=8.5cm,height=6.0cm}}
\caption{Sketch of the experimental set-up for a non-local quantum computer
implemented in a CQED system.
The atoms are trapped inside each cavity (atomic traps are not shown).
The external classical fields allow the manipulation of the atomic states.
We show a brief sketch of the apparatus we intend to use to create
entanglement between the atoms.}
\label{Figure2}
\end{figure}

The experimental set-up we propose in this paper is
schematically shown in Fig.~\ref{Figure2}. Within each of
two spatially separated high-$Q$ millimeter-wave cavities a single Rydberg
atom is trapped. The angular frequency of each cavity mode is supposed to be nearly
resonant with the transition frequency between two
Rydberg levels of the respective atom so that the atoms can be modeled as
two-level systems. Let us call $\ket{g}, \ket{e}$ respectively the ground and the
excited atomic state. Qubits $A$ and $B$ are encoded in the
vacuum and one photon state $\{\ket{0},\ket{1}\}_A$ and
$\{\ket{0},\ket{1}\}_B$ of the two cavity fields while the auxiliary qubits
$\alpha$ and $\beta$ are encoded in the $\ket{g}$, $\ket{e}$ states of the
two atoms. To be more specific, in what follows we can consider 
the two circular levels of rubidium atoms with principal quantum numbers $\mu_e=50$ (for the excited state $\ket{e}$) and $\mu_g=49$ (for the ground state $\ket{g}$) \footnote{Circularity means that the angular quantum number of each considered level is maximum: $l=\mu-1$.}. We neglect here the hyperfine structure of the chosen atomic levels which is hardly resolved in a realistic experiment \cite{Harochemicromaser}. The $\ket{e}\leftrightarrow\ket{g}$ transition frequency is $\nu_{0}\simeq54$ GHz (wavelength $\lambda_{0}\simeq6$ mm). The radiative lifetime $\tau_{atom}$ of these circular levels of the Rydberg spectrum of rubidium is about $30$ msec \cite{haroche1} while, for $Q\simeq10^8$ and a cavity mode frequency $\nu$ nearly resonant with $\nu_{0}$, the field energy damping time $\tau_{field}$ ranges from $1$ up to $30$ msec \cite{haroche1}. Each cavity can be cooled by a refrigerator in order to avoid blackbody radiation. The atom-cavity field coupling factor, measured by the Rabi frequency $\Omega$, can be as large as $10^5$ sec$^{-1}$ so that we can write $1/{\Omega}\ll\tau_{field},\tau_{atom}$. This means that it is possible to observe coherent interaction between atom and cavity field mode before the occurrence of dissipative or decohering effects due to the relaxation of the cavity fields or the decaying of the atoms. Thus, in the following, we neglect any decoherence mechanism and the dynamics of the whole atom-cavity mode system is governed by a Schr\"odinger
equation \cite{haroche1,giovannetti}. In such set-up the interaction of each two-level atom with a single mode of
electromagnetic field is well described by the Jaynes-Cummings Hamiltonian model \cite{jaynescummings}:
\begin{eqnarray}
\label{JaynesCummings}
H_{JC}=\frac{1}{2}\hbar\omega_0{\sigma}_z+\hbar\omega\left(a^{\dagger}a+\frac{1}{2}\right)
+\hbar\Omega\left(a^{\dagger}{\sigma}_{-}+a{\sigma}_{+}\right)
\end{eqnarray}
where $a^{\dag}$ and $a$ are the bosonic operators and $\omega=2\pi\nu$ the angular
frequency of the field mode; $\sigma_{+}=\sigma_{-}^{\dag}=\ket{e}\bra{g}$, $\omega_{0}=2\pi\nu_{0}$ and
$\sigma_{z}=(\ket{e}\bra{e}-\ket{g}\bra{g})$ are
the raising atomic operator, the $\ket{e}\leftrightarrow\ket{g}$ angular
frequency and the third Pauli operator respectively.

In the Hilbert space ${\cal H}={\cal H}_{atom}\otimes{\cal H}_{field}$, the
state $\ket{g,0}$ is an eigenstate
of $H_{JC}$ with energy $E_{g,0}=-\hbar\delta/2$, where
$\delta=\omega_{0}-\omega$ is the atom-cavity mode
detuning. Apart from $\ket{g,0}$, the energy spectrum of the system is
divided into manifolds, essentially labeled by the number of photons
in the cavity field $n$, each formed by the unperturbed states
$\{\ket{e,n},\ket{g,n+1}\}$ (for $n\ge0$). The diagonalization of $H_{JC}$,
in each manifold with an assigned
value of $n$, leads to the well-known dressed states:
\begin{eqnarray}
\label{dressedstates1}
\begin{cases}
\ket{{\cal V}^{n}_{+}}=\cos{\varphi_n}\ket{e,n}+\sin{\varphi_n}\ket{g,n+1}\\
\ket{{\cal
V}^{n}_{-}}=-\sin{\varphi_n}\ket{e,n}+\cos{\varphi_n}\ket{g,n+1}\\
\end{cases}
\end{eqnarray}
where $\tan{\left(2\varphi_n\right)}=(2\Omega\sqrt{n+1})/{\delta}$
\cite{haroche1}. The corresponding eigenenergies are:
\begin{equation}
\label{dressedenergies}
E^{(n)}_{\pm}=\hbar\omega(n+1)\pm\hbar\sqrt{(\delta/2)^2+\Omega^2(n+1)}.
\end{equation}
By suitably varying the detuning $\delta$ it is possible to couple -
decouple the atom and the cavity mode and
to coherently mix the bare states which in the following will be used to
encode quantum information. Assume for
instance  that at $t=0$  the state of the system is $\ket{g,n+1}$ and that
we suddenly switch $\delta=0$: the state will undergo a Rabi flipping as
\begin{equation}
\label{Rabi1}
\ket{\psi(t)}=\cos{\left(\Omega\sqrt{n+1}{t}\right)}\ket{g,n+1}
-i\sin{\left(\Omega\sqrt{n+1}{t}\right)}\ket{e,n}.
\end{equation}

On the contrary, in the {\em dispersive regime} defined by $\Omega\sqrt{n+1}\ll\delta$ the atom is decoupled
from the cavity and there is no coherent exchange of quantum excitations between atom and field and quantum Rabi
oscillations are absent.

We conclude this section by describing how an external classical field couples with the dressed states. Suppose that
an external pulse, sufficiently intense to be considered a classical field, is switched on the atom inside the
cavity. In~\cite{haroche1}, this external source is a high-frequency Schottky diode, able to provide a quasi-monochromatic field tunable between $40$ GHz and $300$ GHz (see~\cite{haroche1} and references within). The shape of the field is mathematically described by a smooth function. In electric dipole
approximation, the Hamiltonian describing the atom-external pulse interaction can be written as
\begin{equation}
\label{ext1}
H_S(t)=\hbar{g(t)}\{{\sigma}_{+}+{\sigma}_{-}\}
\end{equation}
where $g(t)$ is  a function that includes the shape of the pulse and the
atom-field coupling coefficient~\cite{giovannetti}.
It is straightforward to rewrite (\ref{ext1}) in terms of dressed states.
One important point is that the
external field couples dressed states that belong to adjacent manifolds
only: ${H}_{S}(t)$ has non-null matrix
elements just for dressed states that satisfy $\Delta{n}=\pm{1}$. This fact
allows us to extract a simple $4\times4$ block, relative to the subspace spanned by $\left\{\ket{{\cal V}_{+}^{n-1}},\ket{{\cal V}_{-}^{n-1}},
\ket{{\cal V}_{+}^{n}},\ket{{\cal V}_{-}^{n}}\right\}$,
from the matrix representing ${H}_{S}(t)$:

\begin{widetext}
\begin{equation}
\label{interazione}
H_{S}^{(n)}=\hbar{g(t)}
\begin{pmatrix}
0&0&{\cos\varphi_n\sin\varphi_{n-1}}&{-\sin\varphi_n\sin\varphi_{n-1}}\\
0&0&{\cos\varphi_n\cos\varphi_{n-1}}&{-\sin\varphi_n\cos\varphi_{n-1}}\\
{\cos\varphi_n\sin\varphi_{n-1}}&{\cos\varphi_n\cos\varphi_{n-1}}&0&0\\
{-\sin\varphi_n\sin\varphi_{n-1}}&{-\sin\varphi_n\cos\varphi_{n-1}}&0&0
\end{pmatrix}
\end{equation}
\end{widetext}

$H_{S}(t)$ sums to Hamiltonian $H_{JC}$, which is diagonal in the dressed
states basis, and the Schr\"odinger equation for the time evolution of an arbitrary state
$\ket{\xi}=a(t)\ket{{\cal V}_{+}^{n-1}}+b(t)\ket{{\cal V}_{-}^{n-1}}+c(t)
\ket{{\cal V}_{+}^{n}}+d(t)\ket{{\cal V}_{-}^{n}}$ leads to a system of coupled differential
equations with time-dependent coefficients
that, in general, is not easy to solve. We will see that, under precise conditions on
$\delta$ and on the pulse properties, some important
approximations could be performed on these equations. We will show how the interaction regimes described 
briefly above can be used for the realization of a non-local $\cnot$.

\section{Non-Local {\cnot}}
\label{cnotnonlocale}
In this section we describe how to implement, in our CQED systems, the
optimal protocol for a non-local
$\cnot$ operation. In our scheme, the control qubit
of the gate is encoded in the zero
and one photon states of a mode of the electromagnetic field sustained by
cavity $A$. Similarly, cavity $B$
sustains the field mode representing the target qubit. The initial state of
modes $A$ and $B$ will be prepared in
\begin{equation}
\label{initial}
(a\ket{1}+b\ket{0})_A\otimes(c\ket{1}+d\ket{0})_B.
\end{equation}

We want to prove that the experimental scheme
we propose is able to change this state into
\begin{equation}
a\ket{1}_A\otimes(c\ket{0}+d\ket{1})_B+b\ket{0}_A\otimes(c\ket{1}+d\ket{0})
_B.
\end{equation}

We give here the entire list of operations to implement the non-local $\cnot$,
leaving to the following subsections a detailed
treatment of each one.

\begin{enumerate}
\item{\textbf{Trapping}: the two-levels atoms $\alpha$ and $\beta$ should be
trapped inside the spatially
separated  microwave cavities.}

\item{\textbf{Setting of the initial state of the register}: using
$\pi$-Rabi pulses, we prepare the initial
state of modes $A$ and $B$ and of atoms $\alpha$ and $\beta$.}

\item{\textbf{Setting of the ebit: preparation of an entangled atomic
state}. We set entanglement in the joint
state of the trapped atoms letting $\alpha$ and $\beta$ interact directly
with a previously prepared entangled single-photon state.\\
Even if the expression ``single-photon state'' seems to be more appropriate for the visible range of frequency, it will however be used, in what follows, for millimeter-wave radiation too. In this case, we simply want to indicate the state of a field with a single quantum of excitation whose energy, measured in frequency units, falls into the microwave region of the radiation spectrum.} 

\item{\textbf{Local ${\sf CNOT}$ cavity A$\rightarrow$atom
$\alpha$ and measurement of the state
of atom $\alpha$}: the gate is implemented driving, by an external laser
pulse, a transition between two
specific levels of the dressed spectrum of atom $\alpha$. The measurement of
the atomic state is made inducing cyclic transitions to a third level and detecting the subsequent signal with a
{\it millimeter-wave\hskip0.1cm{r}eceiver}.}

\item{\textbf{Local ${\sf CNOT}$ atom $\beta\rightarrow$cavity B}:
we realize this transformation with a two-photon transition between two particular
dressed states of atom $\beta$ and using a ${\sf CNOT}$ cavity $B\rightarrow${a}tom $\beta$.}

\item{\textbf{Hadamard transform on atom $\beta$}: using $\pi/2$-pulses we create linear combinations,
with equal weights, of states $\ket{e}_{\beta}$ and $\ket{g}_{\beta}$.}
\end{enumerate}

\subsection*{Step 1: trapping of the atoms inside the cavities}

We need to trap each atom inside its respective cavity for a time sufficient
to perform every step required by the protocol for a non-local $\cnot$. Furthermore,
the trapping volume should be as small as possible to pledge a strong atom-cavity field coupling.

These features, long trapping time and small volumes, are usually typical of
a Far-Off-Resonance-Trap (FORT)~\cite{fort1}. This is realized by a very focused laser beam of
frequency tuned below the atomic resonance~\cite{phillips}. In these conditions the dipole force confines the
atom in a potential well. Cooling is obtained by means of the scattering force
furnished by optical molasses~\cite{opticalmolasses}. This mixture of dipole and scattering force
characterizes this trap as an hybrid one \cite{phillips}.

Using a FORT to confine neutral atoms is a common practice in the optical domain and allows to reach
trapping time of a hundred of seconds, in a high vacuum environment~\cite{o'hara}. Recently,
trapping a single atom in a cavity using Magneto-Optic-Traps (MOT)~\cite{phillips} and a FORT
has been proved~\cite{fort2}. The trapping times can be improved if a cryostat is used in addition~\cite{o'hara}.

In the microwave range of frequency, on the other hand, the work by Spreeuw {\it et al}.~\cite{spreeuw} proved experimentally the possibility 
to combine an MOT and a microwave cavity. A MOT and a system of optical molasses are there used to load a 
microwave cavity with an ensemble of alkali atoms. The minimum of the MOT trapping potential is located in the center of the cavity and the temperature of the atoms is kept, by the optical molasses, between $3$ and $5$ $\mu$K~\cite{spreeuw}. Even if the experiment has been performed with a large number
of atoms, it however represents an insight into the realistic mixing of microwave cavities and conventional optical trapping techniques (furthermore, an alternative trapping scheme that uses microwaves and an external static magnetic field as a trap for neutral atoms has been addressed both theoretically and experimentally~\cite{spreeuw,agosta}).

The recent and fast improvement of the technique of atomic trapping and the increase in the control of microwave resonators 
allow to consider the {\it scenario} we propose as not far from practical realization. Particularly promising, in this context, are the 
recently developed techniques for the realization of arrays of single, selectively addressable dipole traps that, because of their very reduced dimensions, could be implanted directly inside the cavity without spoiling its $Q$ quality factor too much~\cite{birkl}.   


\subsection*{Step 2: preparation of the distributed register}

The value of the detuning $\delta=0$  can be controlled by means of the
so-called {\em Stark switching
technique} that uses an external electric field applied to the
atom~\cite{brewer}. If the atom has no permanent
electric dipole moment, the Stark effect will be quadratic in the electric
field amplitude~\cite{Weissbluth}.
This induces a relative shift on the atom's energy spectrum between states
$\ket{e}$ and $\ket{g}$ so that the
transition frequency changes from $\omega_{0}$ to another frequency $\omega'_{0}$.
The cavity mode frequency, on the other hand, will
remain unchanged. The width and the amplitude of the Stark field pulses
can be controlled with high
accuracy, allowing a very precise control of the atomic level
separation~\cite{brewer}. Choosing the amplitude
of the Stark field in such a way that
$\omega'_{0}-\omega\gg2\Omega\sqrt{n+1}$, we are able to decouple the atom
from the cavity field. In such dispersive regime it is possible to change
the atomic state without modifying the
cavity mode population by means of an external electromagnetic field,
resonant with the new transition frequency
$\omega'_{0}$. It is therefore possible to prepare qubit
$\alpha$ ($\beta$) in $\tilde{a}\ket{e}_\alpha+\tilde{b}\ket{g}_\alpha$
($\tilde{c}\ket{e}_\beta+\tilde{d}\ket{g}_\beta$).

The coefficients of these linear combinations can be set fixing
the width of each pulse. If the cavities are initially prepared in
$\ket{0}_A\otimes\ket{0}_B$ (a procedure to
obtain such initial states for the cavities is suggested in
ref.~\cite{Raimond}) the joint system
A+B+$\alpha$+$\beta$ will be described by the tensor product state:
\begin{eqnarray}
\ket{\varphi}=
(\tilde{a}\ket{e}+\tilde{b}\ket{g})_\alpha\otimes(\tilde{c}\ket{e}+\tilde{d}
\ket{g})_\beta\otimes\ket{00}_{AB}
\end{eqnarray}

By switching the detuning to its value $\delta=0$ a  coherent
exchange of quantum excitations between
each atom and the relative cavity field takes place. If we leave them to interact
resonantly for a time $t =\pi/(2\Omega)$ ($\pi$-Rabi pulse)
we obtain a complete inversion
between states $\ket{0e}$ and $\ket{1g}$ for each atom+cavity system (see
section~\ref{system}). State
$\ket{0g}$, on the contrary, being an eigenstate of $H_{JC}$, is not
modified by the resonant interaction. The operations described above, based on the atom-cavity mode local resonant interaction, 
can be realized with high accuracy~\cite{haroche1}. The
effect of the $\pi$-pulses is to change $\ket{\varphi}_{\alpha\beta{AB}}$
into:
\begin{equation}
\begin{split}
&\ket{g}_{\alpha}\otimes\ket{g}_{\beta}\otimes\ket{\varphi_{in}}_{A}\otimes
\ket{\varphi_{in}}_{B}=\\
\label{statiiniziali}
&\ket{g}_{\alpha}\otimes\ket{g}_{\beta}\otimes(a\ket{1}+b\ket{0})_A\otimes(c
\ket{1}+d\ket{0})_B.
\end{split}
\end{equation}
We have created initial states of the cavity modes which are coherent
superpositions of Fock states with zero
and one photon. The state for atoms $\alpha+\beta$ is, at this moment,
$\ket{g}_{\alpha}\otimes\ket{g}_{\beta}$: we will
further manipulate it, in the next step of our scheme, to prepare the
required ebit.


\subsection*{Step 3: preparation of the atomic ebit}

In the scheme we propose, the trapped atoms $\alpha$ and $\beta$ encode an ebit, the quantum resource to the
non-local implementation of the $\cnot_{AB}$. The state could be prepared by letting the atoms interact with a pair of external microwave fileds previously prepared in the maximally entangled state: 
\begin{equation}
\label{fotoni}
\ket{\psi_+}=\frac{1}{\sqrt{2}}(\ket{01}+\ket{10}).
\end{equation}

A possible way to get this state of radiation is using a {\it photon\hskip0.1cm{gun}}, 
a device which is able to generate single photon wave packets on demand, in a nearly deterministic way. There are many proposals 
for single-photon sources: semiconductor quantum dots, one-photon emission by isolated molecules, stimulated adiabatic rapid passages of
neutral atoms strongly coupled to a resonator or strongly attenuated beams~\cite{gun}. In this attenuation scheme, a pulsed 
laser field is simply attenuated with density filters until there is on average a fraction of a photon per pulse~\cite{bourennane}. 
The technique should be feasible and can be accomplished even at the range of wavelengths relevant to our set-up. 
The pulse can then be sent to a $50:50$ beam splitter
(BS), whose second input is the vacuum field. The effect of the BS is to mix the two input fields. By properly setting the relative phase between the 
output fields by means of a phase shifter, the joint output state is the maximally entangled state written in Eq.~(\ref{fotoni})~\cite{tanewalls}.

Another possible scheme for the generation of single-photon states of radiation is based on the {\it no-pass} scheme of Hong and Mandel~\cite{Mandel}: via a process of Spontaneous-Parametric-Down-Conversion (SPDC)~\cite{spdc}, a signal and an idler photon are simultaneously generated (within an incertitude of 100 psec~\cite{Mandel}). The two emitted photons are entangled in momentum~\cite{rarity} and if one of them is detected, at some position and within a temporal window $T$, then we are sure that there is a conjugate idler photon, in the corresponding position and inside the same window. Thus, if by means of a photon-counting apparatus a single photon is detected in the signal, the idler is istantaneously prepared in a single-photon state. Furthermore, the experiment performed by Hong and Mandel has shown that the probability that more than just one signal photon is generated by the SPDC is negligible with respect to that of a single-photon generation. This procedure is thus able to create, with a good accuracy, a single-photon state of radiation that can, then, be sent to the same BS described above in order to realize state~(\ref{fotoni}).

The next step we have to perform to get an entangled atomic state is the direct interaction of the trapped
atoms $\alpha$ and $\beta$ with $\ket{\psi_+}$. The entangled photons we
prepared are sent to the
atoms (via suitable designed waveguides directly coupled to the cavities $A$ and $B$, for example). In each cavity-atom subsystem,
a dispersive regime of interaction should be set, so that the atom dynamics
is decoupled from that of the cavity.

If the spectrum of each light pulse is sufficiently narrow and centered at a frequency resonant to the atomic transition
$\ket{g}\leftrightarrow\ket{e}$, it is possible to show that setting the interaction time in order to realize a $\pi$-pulse, the following transformation can be realized:
\begin{equation}
\ket{g}_{\alpha}\otimes\ket{g}_{\beta}\rightarrow\frac{1}{\sqrt{2}}(\ket{eg}
+\ket{ge})_{\alpha\beta}.
\end{equation}

However, these two techniques are not immune to problems. In the case of the attenuated beam, at a so 
low intensity level, it is not possible to be sure there being a photon. There is always 
a possibility to get an empty pulse or a two-photon one. In the latter case, even 
if the procedure described is able to generate atomic entanglement, the state consequently obtained 
is not of the form we need. Even more, as we have seen, a precise control of the area of the pulse is required in order to accomplish exactly the required atomic evolution.

On the other hand, the Hong and Mandel scheme presents some difficulties for the microwave range of frequencies because it is based on a SPDC process. The generation efficiency of the couple of conjugate photons, in a down-conversion process, is very low (a rough but, for our purposes, sufficient semiclassical approach to the theory of SPDC shows that it is directly proportional to the fourth power of the pump beam frequency $\omega^4_p$). Tipically, optical frequency beams are used as pumps for SPDC (Ultra-Violet in~\cite{Mandel}) and this gives a rate of generation of down converted photons of the order of $10^{-10}$ sec$^{-1}$. For microwave frequencies the rate of down conversion is dramatically smaller than this value and makes the scheme useless as a photon gun. Even more, technical difficulties, in the microwave case, have to be managed. The crystal used for the generation of signal and idler, for example, has to be almost transparent to the frequency of both the pump and the down converted fields. Finding an appropriate candidate for a pump that falls in the microwave range is a non trivial task.
\begin{figure}[ht]
\centerline{\psfig{figure=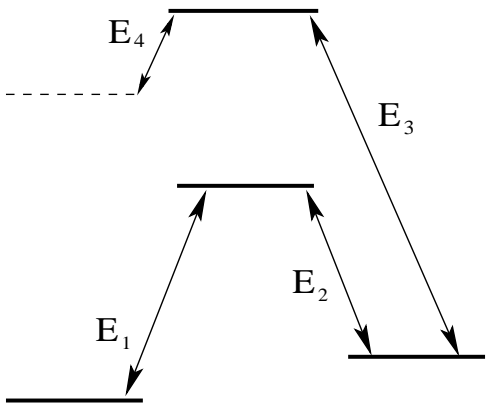,width=6.5cm,height=5.3cm}}
\caption{Scheme of the atomic levels for the generation of a microwave field by means of a four-
wave-mixing process. If $E_2$ is slowly turned off while $E_1$ enters in an optically thick 
medium, the quantum information carried by $E_1$ is stored in the coherences established between the ground, metastable states of the atomic model.
 If the {\it reading} field $E_3$ is then 
turned on, for properly chosen values of its frequency, a fourth field, in the microwave range 
of frequencies, could be generated.}
\label{lambda}
\end{figure}

A deterministic source of one-photon states has been theoretically proposed and, recently, experimentally 
demonstrated in the microwave range of frequencies~\cite{walther}: it is based on the interaction of a beam of Rydberg atoms with a field mode of a very high-$Q$ microwave cavity (in the experiment $Q\sim10^{10}$). Unfortunately, the enormous $Q$ value of the cavity used in~\cite{walther} prevents significative leakage of the field from the cavity itself and, thus, the scheme does not produce any exploitable output beam.

The recently developed techniques to store the quantum state of a field in a macroscopic atomic ensemble that exibits electromagnetically-induced-transparency (EIT)~\cite{lukin} could represent a possible solution to the problem represented by the production of a microwave single photon state. Assume that the preparation of a one-photon state of an optical pulse, as experimentally done in~\cite{Mandel}, is followed by a {\it storage} step in an optically thick medium made by three-level atoms in a $\Lambda$ configuration~\cite{lukin}. For the notation of the following discussion we refer to Fig.~\ref{lambda}. If the field $E_2$, with wavevector ${\bf k}_2$, is slowly turned off while $E_1$ (wavevector ${\bf k}_1$) interacts with the atomic medium, the quantum information about the amplitude, shape and phase of the latter is transferred to the coherences established between the two ground (metastable) states of the atomic model. To recover the information so stored, a {\it reading} pulse is necessary. If a field $E_3$ is shined on the ensemble, the new electromagnetic field $E_4$, with a wavevector ${\bf k}_4$ that satisfies the {\it phase matching condition} ${\bf k}_1+{\bf k}_2={\bf k}_3+{\bf k}_4$, is generated by means of a process of forward four-wave-mixing~\cite{zibrov}. Properly choosing the values for the angular frequencies $\omega_{j}=ck_j\hskip0.2cm(j=1,3)$, the generated field can fall in the microwave range~\cite{ham}.  


\subsection*{Step 4: local $\sf CNOT$ $A\rightarrow\alpha$ and measurement
of the state of atom $\alpha$}

The theoretical protocol requires a $\cnot$ having cavity $A$ as control
and atom $\alpha$ as target. Since
this unitary operation involves just one cavity and the respective trapped
atom, we refer to $\cnot_{A\alpha}$ as a local transformation to distinguish it from the non-local one we want to perform between
cavity $A$ and cavity $B$. The computational basis for the $\cnot_{A\alpha}$ is
$\{\ket{0g},\ket{0e},\ket{1g},\ket{1e}\}_{A\alpha}$ and the set of
transformations we should realize is:
\begin{equation}
\label{cnotteorica}
\begin{split}
\ket{0g}_{A\alpha}&\rightarrow\ket{0g}_{A\alpha}\hskip0.5cm
\ket{0e}_{A\alpha}\rightarrow\ket{0e}_{A\alpha}\\
\ket{1g}_{A\alpha}&\rightarrow\ket{1e}_{A\alpha}\hskip0.5cm
\ket{1e}_{A\alpha}\rightarrow\ket{1g}_{A\alpha}.
\end{split}
\end{equation}

According to this set of transformations, the state prepared during Step 3:
\begin{equation}
\ket{\chi}=\frac{1}{\sqrt{2}}(\ket{eg}+\ket{ge})_{\alpha\beta}(a\ket{1}+b\ket{0})_{A}(c\ket{1}+d\ket{0})_{B}
\end{equation}
has to be changed into:
\begin{equation}
\label{teoria}
\begin{split}
\cnot_{A\alpha}\ket{\chi}&=\frac{1}{\sqrt{2}}\left\{a\ket{1}_{A}(\ket{gg}+\ket{ee})_{\alpha\beta}\right.\\
&\left.+b\ket{0}_{A}(\ket{eg}+\ket{ge})_{\alpha\beta}\right\}(c\ket{1}+d\ket{0})_{B}.
\end{split}
\end{equation}

Expressions~(\ref{cnotteorica}) and~(\ref{teoria}) show that, while the atomic state can modify its state, the cavity mode
population does not change: this means that a resonant coupling between $A$ and $\alpha$ can not be used to
implement the gate. Resonant Rabi oscillations, indeed, preserves the total number of excitations while the last
two transformations in (\ref{cnotteorica}) do not. We need a dispersive atom-cavity field interaction; the
atomic state will be manipulated by an external electromagnetic field. If the external field is resonant with a field mode sustained by the cavity but different from the relevant one used to codify the cavity qubit, it can enter the resonator without feeding this latter mode (for example we can choose two orthogonally polarized field modes: in~\cite{haroche1} two orthogonally polarized transverse modes, with a spacing in frequency of 70 kHz, sustained by a millimeter-wave cavity are considered). Using the Stark switching technique, the trapped atom can then interact with the external field, being decoupled with respect to the relevant cavity mode, for a controlled amount of time.

The Stark field can be set to change the separation between levels
$\ket{e}_{\alpha}$ and $\ket{g}_{\alpha}$
and to obtain a value of $\delta$ that allows to write
$\Omega\ll\delta$. In such a condition, from Eq.~(\ref{dressedstates1}),
it results that $\ket{{\cal V}^{n}_+}\simeq\ket{n,e},\ket{{\cal V}^{n}_-}\simeq\ket{n+1,g}$. Therefore, if using an appropriate
external pulse we can induce a complete inversion of population between
$\ket{{\cal V}^{1}_+}$ and
$\ket{{\cal V}^{0}_-}$, we mutually exchange $\ket{1e}$ and $\ket{1g}$
without involving the other dressed states,
$\ket{{\cal V}^{0}_+}\simeq\ket{0e}$ and $\ket{{\cal
V}^{1}_-}\simeq\ket{2g}$. The only approximative identification of the bare basis elements with the corresponding dressed states in the limit of large detuning has only a very small effect on the fidelity of the gate, as we show below.

Since we want the interaction of atom $\alpha$ with an external pulse, we
use results obtained in \ref{system}:
the dressed manifolds involved in the transition $\ket{{\cal
V}^{1}_+}\leftrightarrow\ket{{\cal V}^{0}_-}$
are those with $n=0$ and $n=1$ (therefore satisfying condition
$\Delta{n}=\pm1$). We give explicit expression
for $H_{S}^{(1)}(t)$ expanding each matrix element in Taylor series, to
second order in $\Omega/\delta$:

\begin{equation}
\label{interazione2}
\frac{H_{S}^{(1)}}{\hbar}={g(t)}
\begin{pmatrix}
0&0&{\frac{\Omega}{\delta}}&{-\sqrt{2}{\frac{\Omega^2}{\delta^2}}}\\
0&0&{\left(1-\frac{3}{2}{\frac{\Omega^2}{\delta^2}}\right)}&{-\sqrt{2}\frac{
\Omega}{\delta}}\\
{{\frac{\Omega}{\delta}}}&{\left(1-\frac{3}{2}{\frac{\Omega^2}{\delta^2}}\right)}&0&0\\
{-\sqrt{2}{\frac{\Omega^2}{\delta^2}}}&{-\sqrt{2}\frac{\Omega}{\delta}}&0&0
\end{pmatrix}
\end{equation}

We deduce that if the experimental parameters are such that
$\Omega\ll\delta$, the matrix elements belonging to the central $2\times2$
block in $H_{S}^{(1)}$ predominates over all the others.
We choose $g(t)$ so that $g(t)=p(t)\cos{(\omega_St)}$, with $p(t)$ a smooth
function describing the envelope shape of the pulse shined on the trapped
atom. We set $\omega_{S}$ to be equal to the transition frequency for $\ket{{\cal V}^{1}_+}\leftrightarrow\ket{{\cal V}^{0}_-}$:
\begin{equation}
\omega_{S}=\frac{E_{+}^{1}-E_{-}^{0}}{\hbar}
\end{equation}

With a suitable choice of the pulse duration, the right inversion of population can be obtained. In particular, it has to be $\int^t{p}(t')dt'=\pi$ (we refer to this case as to a {\it{$\pi$ pulse}}).
Any spurious phase factor can be adjusted by setting an appropriate phase in function $g(t)$ or using
appropriate Stark field and, in what follows, we do not care about it~\cite{giovannetti}. Achievable coupling strength for the atom-external field interaction as large as $20\pi$ kHz and interaction times of about $50$ $\mu$sec allow to get a complete $\pi$ pulse. 

The effect of a finite, non null, value of the ratio $\Omega/\delta$ on the state that we instead obtain, can be seen if
we propagate $\ket{\chi}$ by means of the unitary operator that is generated by the Hermitian interaction Hamiltonian $H^{(1)}_{S}(t)$. We assume $x=\Omega/\delta=0.1$, value that allows for the approximations discussed above and for the discrimination between the frequencies of the different transitions involved; retaining just the terms up to the second order in $\Omega/\delta$, we get an approximate expression for the evolved state of the system 
$\ket{\tilde{\chi}}$. This expression is useful in order to find the fidelity of the local $\cnot$ operation we are performing.

The definition of the fidelity function, in this case, reads ${\cal{F}}=\modul{\sand{\chi}{\cnot_{A\alpha}}{\tilde\chi}}^2$~\cite{Preskill} and assuming for simplicity $a=b=\frac{1}{\sqrt{2}}$, it is possible to show that:

\begin{figure}[b]
\centerline{\psfig{figure=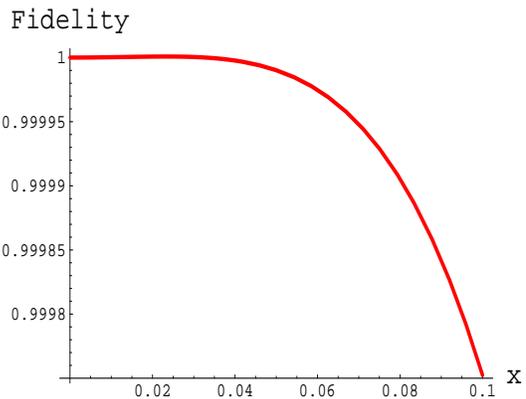,width=7.0cm,height=5.4cm}}
\caption{Plot of the fidelity function ${\cal{F}}$ of the local $\cnot_{A\alpha}$ operation as a function of the ratio $x=\Omega/\delta$ and 
for $a=b=1/\sqrt{2}$. As it is shown, for values of $x$ that range from 0 to 0.1, the gate can be performed with a very high accuracy. 
The high fidelity of the gate is maintained even for different choices of $a$ and $b$.} 
\label{Figure3}
\end{figure}

\begin{equation}
{\cal{F}}(x)=\frac{1}{4}\left\{1+\sin\left[\frac{\pi}{2}\left(1-\frac{3}{2}x^2
\right)\right]\right\}^2+0.003x^2.
\end{equation}
Notice that, as $\Omega/\delta\rightarrow{0}$, the fidelity reaches ${\cal F}=1$, which shows a perfect overlap between the state we need and the one we get manipulating the atomic qubit by an external field. 

The state of the system at the end of Step 3 becomes:
\begin{equation}
\begin{split}
&\frac{1}{\sqrt{2}}\ket{g}_{\alpha}\otimes\left(-ia\ket{1g}+b\ket{0e}\right)_{A\beta}\otimes
\ket{\varphi_{in}}_B+\\
&\frac{1}{\sqrt{2}}\ket{e}_{\alpha}\otimes\left(-ia\ket{1e}+b\ket{0g}\right)_{A\beta}\otimes
\ket{\varphi_{in}}_B.
\end{split}
\end{equation}

We now measure the state of atom $\alpha$ on the bare eigenstate basis $\left\{\ket{e},\ket{g}\right\}_{\alpha}$. We enlarge the Hilbert space of the atom introducing a third energy level, $\ket{m}_{\alpha}$, whose parity is opposite to that of 
$\ket{e}_{\alpha}$. For example, we can take the Rydberg level with principal quantum number $\mu_{m}=51$. 
In this case the $\ket{e}_{\alpha}\leftrightarrow\ket{m}_{\alpha}$ frequency is about $\nu_{m}\simeq51.1$ GHz \cite{haroche1}. 
An external microwave field couples $\ket{e}_{\alpha}$ to $\ket{m}_{\alpha}$. 
If the state of $\alpha$ is $\ket{e}_\alpha$, the external field induces cyclic transitions between these states but, if atom $\alpha$ is in 
$\ket{g}_\alpha$, because of the large frequency mismatch, we do not detect any signal. If $\ket{m}_{\alpha}$ corresponds to a 
low angular quantum number, the emission time of the atom falls in the range of $\mu$sec \cite{haroche1,moi}. The radiation emitted by 
the cycling atom can be collected by a ${\it millimeter-wave~receiver}$~\cite{moi}. 
This is, essentially, a Schottky diode detector that mixes the signal to be measured with a local reference microwave field to perform a heterodyne 
measurement of the signal. The response time of the device is short enough not to represent a limitation for our purposes. This detection technique has been successfully used in the context of micromaser spectroscopy  
to directly infer the radiation of a millimeter-wave field inside a cavity~\cite{haroche1,moi}. 

Depending on the state of atom $\alpha$ at the end of the measurement process, system $A+\beta+B$ is projected onto states 
which differ just for the state of atom $\beta$. In order to obtain the right final state at the end of the protocol for the
non-local $\cnot_{AB}$, if the atomic state detection gives
$\ket{g}_\alpha$, we should change nothing in subsystem $B+\beta$. If the
output of the measurement is $\ket{e}_\alpha$, a $\sf NOT$ is required for
qubit $\beta$.

To obtain  it we essentially need the same kind of transformations we introduced in the last step:
$\ket{1e}_{B\beta}\rightarrow\ket{1g}_{B\beta}$ and $\ket{0g}_{B\beta}\rightarrow\ket{0e}_{B\beta}$. They can be
realized by applying $\pi$-pulses for transitions between suitable dressed states of the atom $\beta$ in a
dispersive regime of interaction. In any case, with a fidelity that approaches $100\%$, the  system $A+\beta+B$ can be set in
$\left(-ia\ket{1g}+b\ket{0e}\right)_{A\beta}\otimes\ket{\varphi_{in}}_B$ apart from a  global phase factor.\\
For simplicity, in the following, we assume to have detected $\ket{g}_{\alpha}$.

\subsection*{Step 5: local ${\sf CNOT}\hskip0.2cm\beta\rightarrow{B}$}

The next step is the implementation of a local $\cnot_{\beta{B}}$.
The atomic qubit $\beta$ is now the control of the gate. To get the right
final state for the non-local $\cnot_{AB}$, the set of transformations to realize
is the following:
\begin{equation}
\label{cnotatomocavita}
\begin{split}
&\ket{g0}_{\beta{B}}\rightarrow\ket{g1}_{\beta{B}}\hskip0.5cm
\ket{g1}_{\beta{B}}\rightarrow\ket{g0}_{\beta{B}}\\
&\ket{e0}_{\beta{B}}\rightarrow\ket{e0}_{\beta{B}}\hskip0.5cm
\ket{e1}_{\beta{B}}\rightarrow\ket{e1}_{\beta{B}}.
\end{split}
\end{equation}
It is clear that, in~(\ref{cnotatomocavita}), we have $\ket{g}\equiv\ket{1}$
and $\ket{e}\equiv\ket{0}$.\\
In this subsection we show how to realize these transformations using
the two-photon transition
$\ket{g0}_{\beta{B}}\leftrightarrow\ket{{\cal V}_{+}^{1}}_{\beta{B}}$ and a
$\cnot_{B\beta}$.

We need such a different strategy because an approach similar to that used for the $\cnot_{A\alpha}$ will lead us to some
inconsistencies. In effect, using the same logic scheme used to implement the
Step 4, a procedure to get transitions (\ref{cnotatomocavita}) could be the following.
We induce a $\pi$-pulse between $\ket{g0}_{\beta{B}}$ and
$\ket{{\cal V}_{-}^0}_{\beta{B}}$. State
$\ket{{\cal V}_{-}^0}_{\beta{B}}$, for $\Omega\ll\delta$, is in practice
the bare state
$\ket{g1}_{\beta{B}}$ but, for the selection rules relative to electric
dipole transitions~\cite{Weissbluth},
transition $\ket{g0}_{\beta{B}}\leftrightarrow\ket{g1}_{\beta{B}}$ is
strictly forbidden. Since $\ket{{\cal V}_{-}^0}_{\beta{B}}$ holds a little contribution from
$\ket{e0}_{\beta{B}}$ \cite{giovannetti}, transition $\ket{{\cal V}_{-}^0}_{\beta{B}}\leftrightarrow\ket{g0}$ can be realized.
Using the same procedure of previous sections we can obtain the matrix
representation of the Hamiltonian
$H_{S}(t)$ for the interaction of the dressed atom $\beta$ with the external
pulse driving the required
transition. On the ordered dressed basis
$\left\{\ket{g0}_{\beta{B}},\ket{{\cal V}_{-}^0}_{\beta{B}},\ket{{\cal V}_{+}^0}_{\beta{B}}\right\}$, we have
\begin{eqnarray}
H_{S}^{(0)}(t)=\hbar{g(t)}
\begin{pmatrix}
0&-\frac{\Omega}{\delta}&{\left(1-\frac{1}{2}{\frac{\Omega^2}{\delta^2}}\right)}\\
-\frac{\Omega}{\delta}&0&0\\
{\left(1-\frac{1}{2}{\frac{\Omega^2}{\delta^2}}\right)}&0&0
\end{pmatrix}
\end{eqnarray}

Here, the biggest matrix elements are those that connect
$\ket{g0}_{\beta{B}}$ to $\ket{{\cal V}_{+}^{0}}_{\beta{B}}$ and vice versa.
For $\Omega\ll\delta$, $\ket{{\cal V}_{+}^{0}}_{\beta{B}}$ is essentially
identified with $\ket{e0}_{\beta{B}}$ \footnote{This explains how to realize
an efficient $\pi$-pulse that could exchange $\ket{0g}_{B\beta}$ and
$\ket{0e}_{B\beta}$ as required, on the state of system $A+B+\beta$, after a
measurement having $\ket{e}_{\alpha}$ as output.}. On the contrary, the
probability of a transition
$\ket{g0}_{\beta{B}}\leftrightarrow\ket{{\cal V}_{-}^{0}}_{\beta{B}}$ is
directly proportional to $\Omega/\delta$. However, since we want
$\Omega\ll\delta$, we need a different procedure.

\begin{figure}[ht]
\centerline{\psfig{figure=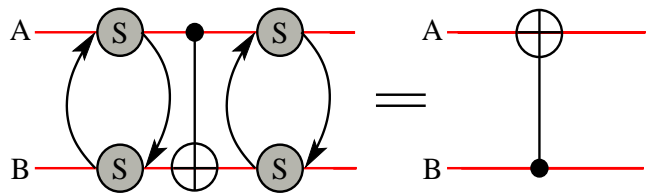,width=8.5cm,height=2.5cm}}
\caption{A $\cnot_{BA}$ gate can
be simulated using the sequence of operations
$({\sf SWAP})(\cnot_{AB})({\sf SWAP})$.} 
\label{Figure4}
\end{figure}

We now proceed mapping the $\cnot_{\beta{B}}$ into a sequence of three
operations: two $\sf SWAP$ and a $\sf CNOT_{B\beta}$ (Fig.~\ref{Figure4}).

It is straightforward to prove that the sequence $({\sf SWAP})(\cnot_{AB})({\sf SWAP})$
is equivalent to $\cnot_{BA}$. Since we have already seen an efficient way to implement
a $\cnot$ for the cavity qubit as control and the atomic one as target, we
just pass to describe a possible procedure to accomplish a $\sf SWAP$
operation in a CQED system.

By the action of the $\gate{SWAP}$ gate, a transitions occurs between
$\ket{g0}_{\beta{B}}$ and $\ket{e1}_{\beta{B}}$. For an
external Stark field satisfying $\Omega\ll\delta$, the dressed state
$\ket{{\cal V}_{+}^{1}}_{B\beta}$ is about equal
to $\ket{1e}_{B\beta}$ as it can be deduced from Eq.~(\ref{dressedstates1}).
Inducing $\ket{{\cal V}_{+}^{1}}_{{B}\beta}\leftrightarrow\ket{0g}_{{B}\beta}$ we
can get what we want. Nevertheless, these dressed states belong to
dressed manifolds which differ for two quantum excitations and that cannot
be connected by a single photon transition. This means that we should go to the
second order in the coupling coefficient, realizing a two-photon transition.
We choose it to be degenerate: this means that the photons involved in this second order
process have the same energy $\hbar\omega_L=\frac{1}{2}(E_{+}^{(1)}-E_{0g})$, where the
external field has frequency $\omega_L$ \cite{haroche3}.

The second order transition $\ket{{\cal V}_{+}^{1}}_{{B}\beta}\leftrightarrow\ket{0g}_{{B}\beta}$ occurs via virtual
transitions toward the intermediary states $\ket{{\cal V}_{\pm}^{0}}_{B\beta}$. In effect, because of the structure of the energy
spectrum of the dressed atom, these states have energies which are very
close to the middle energy between $\ket{0g}_{B\beta}$ and $\ket{{\cal V}_{+}^{1}}_{B\beta}$. However, the transitions are not resonant, 
so  that the probabilitythat the system can accomplish an effective transition to $\ket{{\cal V}_{\pm}^{0}}_{B\beta}$
is negligible. This qualifies $\ket{{\cal V}_{\pm}^{0}}_{B\beta}$ as virtual states.

We model the external field as a linearly polarized pulse of Gaussian
envelope:
\begin{equation}
E(t)={\cal E}_0e^{-\frac{t^2}{\tau^2}}\cos{(\omega_L{t})}.
\end{equation}
The electric dipole interaction gives rise to the following interaction
Hamiltonian:
\begin{equation}
{H}_{L}(t)=\hbar\Sigma_0\left\{e^{-\frac{t^2}{\tau^2}+i\omega_L{t}}\pro{g}{e
}+h.c.\right\}
\end{equation}
where $\Sigma_0$ is the atom-field coupling coefficient and RWA has been
used. The probability amplitude that the system, initially in
$\ket{0g}_{B\beta}$, is found at time $t$ in $\ket{{\cal
V}_{+}^{1}}_{{B}\beta}$ can be calculated using the following expression,
directly derived from a second-order perturbation
theory~\cite{Cohen,photonsandatoms}:
\begin{equation}
\label{probamp}
\begin{split}
&\frac{1}{\hbar^2}\sum_{j=-,+}\left(\int_{-\infty}^{t}dt''\sand{{\cal V}_{+}^{1}}{H_{L}(t'')}
{{\cal V}_{j}^{0}}e^{\frac{i}{\hbar}\left(E_{+}^{(1)}-E_{j}^{(0)}\right)t''}\times\right.\\
&\times\left.\int_{-\infty}^{t''}dt'\sand{{\cal V}_{j}^{0}}{H_{L}(t')}{0g}e^{\frac{i}{\hbar}\left(E_{j}^{(0)}-E_{0g}\right)t'}\right)
\end{split}
\end{equation}
(with $t''>t'$). Explicit evaluation of this expression, with numerical
values $\Omega\approx{10^5}$ Hz, $\delta\approx{1}$ MHz, $\tau\approx{20}$ $\mu${sec},
$\Sigma_{0}\approx{10^5}$ Hz and for $t=3\tau$ leads to a transition probability equal to 0.47.
The value of $\delta$ satisfies the condition $\delta\ll\omega,\omega_0$ because,
for a millimeter-wave cavity, the value of $\omega$ falls in the range from $10$ to $100$~{GHz}
while a typical value for $\omega_{0}$, for values of the principal quantum number $\mu\simeq50$,
is $50$ GHz~\cite{haroche1}. Having
$\Sigma_0\simeq\Omega_0$ ensures the observability of multiphotons
transitions.\\
Since the explicit calculation for the
$\ket{1e}_{B\beta}\rightarrow\ket{0g}_{B\beta}$ case leads, with the same
numerical values of the previous case, to the same probability of
transition, our map of a $\cnot_{\beta{B}}$
is valid for each initial state of the system $\beta+B$. Ideally, we are
able to implement a $\sf
CNOT_{\beta{B}}$ using just ${\sf SWAP}$ operations and $\cnot_{cavity-atom}$.
In order to evaluate the fidelity of the local $\cnot_{\beta{B}}$, we have to perform essentially the same kind of calculation 
described in Step 4 for the case of the local $\cnot_{A\alpha}$ gate. We found that, at the end of the operations, the state of 
the system $A+B+\beta$ is projected onto
\begin{equation}
ia\ket{g}_{\beta}\otimes\ket{1}_{A}\otimes\left\{{\sf NOT}\ket{\varphi_{in}}_{B}\right\}+b\ket{0}_{A}\otimes
\ket{e}_{\beta}\otimes\ket{\varphi_{in}}_{B}
\end{equation}
with a fidelity $0.54$, for $c=d=\frac{1}{\sqrt{2}}$ and after an average over all the possible initial configurations of 
the system $B+\beta$. This low value of the fidelity of the gate is essentially due to the non ideality  
of the two-photon transition and represents the major theoretical limitation to the efficiency of the proposed implementation. 


\subsection*{Step 6: Hadamard transform on atom $\beta$}

We now need a Hadamard transform on qubit $\beta$. This can be realized, in a CQED system,
recurring again to a dispersive regime of atom-cavity field interaction.
In effect, setting a very large detuning (leaving the eigenstates of the atom $\beta$ almost bare)
and shining a driving external pulse on
$\beta$ for a time such that a {$\pi/2$}-pulse is realized between
$\ket{ej}$ and $\ket{gj}$ ($j=0,1$), we just obtain the following transitions:
\begin{equation}
\label{Hadamatom}
\begin{split}
&\ket{gj}\rightarrow\frac{1}{\sqrt{2}}\left\{\ket{gj}-i\ket{ej}\right\}\\
&\ket{ej}\rightarrow\frac{1}{\sqrt{2}}\left\{\ket{ej}-i\ket{gj}\right\}.
\end{split}
\end{equation}

We have, in practice, a Hadamard transform generalized by a relative phase
factor that is not a problem for our scheme: using relations
(\ref{Hadamatom}) in the state obtained at the end of Step 5, we have
\begin{equation}
\begin{split}
&\frac{i}{\sqrt2}\ket{g}_{\beta}\otimes\left\{a\ket{1}\otimes({\sf
NOT}\ket{\varphi_{in}})-b\ket{0}\otimes\ket{\varphi_{in}}\right\}_{AB}+\\
&\frac{1}{\sqrt2}\ket{e}_{\beta}\otimes\left\{a\ket{1}\otimes({\sf
NOT}\ket{\varphi_{in}})+b\ket{0}\otimes\ket{\varphi_{in}}\right\}_{AB}.
\end{split}
\end{equation}

If the measurement outcome of the atom $\beta$ is $\ket{e}_{\beta}$,
system $A+B$ is projected onto a state that shows the
action of the $\cnot_{AB}$ gate. An ulterior manipulation is required if
the measurement outcome is $\ket{g}_{\beta}$. In this case, if we want to
correct the $-1$ relative phase factor that appears in the $A+B$ state we
just perform a $2\pi$ resonant Rabi pulse in the subsystem $A+\alpha$
(we remind that our measurement process is a non demolition one and that
atom $\alpha$ can always be forced to occupy state $\ket{g}_{\alpha}$, as we
assumed all along our discussion).

This closes the scheme for a non-local $\sf CNOT$ between two spatially
separated cavity modes.

\section{Non-Local {$\pi-\sf CQPG$}}
\label{qpgnonlocale}

In this section we will describe a procedure for the physical implementation
of a non-local controlled quantum phase gate with $\phi=\pi$. This is a very important task 
to accomplish because the set of quantum gates that comprehends controlled quantum phase gate and 
single qubit rotations is adequate for quantum computation. Our goal is to show that the experimental 
set-up proposed in this paper is sufficiently flexible to permit, with slight modifications operated 
in Step 5 of the previous protocol, its feasible realization. As before, the computational
register is formed by the two spatially separated cavity modes $A$ and $B$
while the atoms $\alpha$ and $\beta$ encode two ancillary qubits whose joint
state constitutes an ebit.

We assume that the initial state for system $A+B$ has been prepared as in Eq.~(\ref{initial}).
Moreover, we assume to have a maximally entangled atomic ebit. We want to show how
to transform state $\ket{\varphi_{in}}_{A}\otimes\ket{\varphi_{in}}_{B}$ into
\begin{equation}
ac\ket{11}_{AB}+ad\ket{10}_{AB}+bc\ket{01}_{AB}-bd\ket{00}_{AB}.
\end{equation}

The experimental scheme for the non-local {\sf $\pi$-CQPG} is identical until Step 4 to
that for the non-local $\cnot_{AB}$. This means that, at the end of Step 4, having
performed the local $\sf CNOT_{A\alpha}$ and measured the state of atom $\alpha$, the state of system
$A+B+\beta$ is projected onto
\begin{equation}
\label{dopocnot}
\left(-iac\ket{1g1}-iad\ket{1g0}+bc\ket{0e1}+bd\ket{0e0}\right)_{A\beta{B}}
\end{equation}
while atom $\alpha$ is assumed to be in $\ket{g}_{\alpha}$.

We modify the previous scheme replacing Step 5 with the following to
perform a $\pi-{\sf CQPG}$ on system $\beta+B$. We adopt the following map of the $\sf CQPG$:
\begin{equation}
\label{phasegate}
\begin{split}
&\ket{e0}\rightarrow\ket{e0}\hskip0.5cm\ket{e1}\rightarrow-\ket{e1}\\
&\ket{g0}\rightarrow\ket{g0}\hskip0.5cm\ket{g1}\rightarrow\ket{g1}.
\end{split}
\end{equation}
This set of transformation is obtained by extending the atomic model to comprehend a
third energy level. We introduce state $\ket{i}_{\beta}$ whose parity is opposite to that
of $\ket{e}_{\beta}$. For example, as we did above, a possible choice for the atomic levels can be 
$\ket{i}\leftrightarrow\mu_{i}=51$, $\ket{e}\leftrightarrow\mu_{e}=50$, $\ket{g}\leftrightarrow\mu_{g}=49$.

If we set the cavity mode $B$ resonant to $\ket{e}\leftrightarrow\ket{i}$ the cavity field
results out of resonance with $\ket{g}\leftrightarrow\ket{e}$ and
transitions at this frequency are strongly suppressed. Thus, setting
resonance between atom $\beta$ and cavity mode $B$ for a time sufficient to realize a $2\pi$-Rabi pulse between
$\ket{e1}_{\beta{B}}$ and $\ket{i0}_{\beta{B}}$, state $\ket{e1}_{\beta{B}}$
will acquire a $-1$ phase factor \cite{giovannetti, rauschen}. The phases of the other
states that appear in~(\ref{dopocnot}) are not modified: the phases of $\ket{g0}_{\beta{B}}$ and
$\ket{g1}_{\beta{B}}$ are unchanged because of the frequency mismatch while that of $\ket{e0}$ does not change
because it is an eigenstate of the Jaynes-Cummings Hamiltonian in the bidimensional Hilbert space 
spanned by $\left\{\ket{e},\ket{i}\right\}_{\beta}$ . We are,
thus, able to perform transformations (\ref{phasegate}) and the state in Eq.~(\ref{dopocnot}) changes into:
\begin{equation}
ia\ket{g}_{\beta}\otimes\left(c\ket{11}+d\ket{10}\right)_{AB}+b\ket{e}_{\beta}\otimes\left(c\ket{01}-d\ket{00}\right)_{AB}
\end{equation}
apart from a global phase factor.

Now, we need the set of transformations, on the system $\beta+B$,
defined in Step 6 of the previous protocol. This make us obtain the
final state:
\begin{equation}
\begin{split}
i&\ket{g}_{\beta}\otimes\left(ac\ket{11}+ad\ket{10}-bc\ket{01}+bd\ket{00}\right)_{AB}+\\
&\ket{e}_{\beta}\otimes\left(ac\ket{11}+ad\ket{10}+bc\ket{01}-bd\ket{00}\right)_{AB}.
\end{split}
\end{equation}

If the measurement outcome of atom $\beta$ is $\ket{e}_{\beta}$, the joint state of the
two cavities is such that the action of the $\pi-{\gate{ CQPG}}$ on qubits $A$ and $B$ is evident.
If the outcome of the measurement is $\ket{g}_{\beta}$, we apply a $2\pi$-Rabi pulse for transition
$\ket{g1}_{\alpha{A}}\leftrightarrow\ket{e0}_{\alpha{A}}$ on system $\alpha+A$. It is relevant to notice that, 
in order to realize the local quantum phase gate of the protocol, just simple resonant Rabi oscillation in a atom-cavity system is required. 

We have proposed a non-local $\pi-\gate{CQPG}$ between spatially separated cavities.
Implementing a non-local $\gate{CQPG}$ is an important result, in quantum distributed
computation, since it can help us in improving the efficiency of the non-local $\sf CNOT$.
The non ideality of Step 5 of the scheme for the $\cnot_{AB}$ strongly limits the
efficiency of the gate. Since a $\cnot$ operation can be simulated using a Hadamard transform
on the target qubit followed by a $\pi-{\gate{ CQPG}}$,
and the efficiency of implementation of a non-local $\pi-{\sf CQPG}$
is evidently better than that of the
$\cnot$, the reliability of the non-local gate can be significantly
improved. 

In this case, however, the time needed to accomplish the entire non-local $\cnot$ can represent a problem. The realization of a Hadamard 
transform of the cavity field requires a map of the quantum state of the field onto the relative ancillary (atomic) qubit. The performance of a 
Hadamard transform on the latter and, eventually, a map of the transformed state back onto the cavity qubit. This increases the time necessary to 
accomplish the non-local $\cnot_{AB}$ and the number of local transformations involved.


\section{conclusions}

In this paper we have proposed a CQED set-up for the implementation of a non-local
$\sf CNOT$ and of a non-local $\pi-{\gate{CQPG}}$. According to the
optimal theoretical protocols described in ref. \cite{Plenio}, our
experimental schemes use just two bits of classical communication and a
single ebit, shared by the two parties.

The computational register, in the proposed set-up, is formed by two
spatially separated microwave cavities while the required ebit is encoded in
the entangled state of two Rydberg atoms. 

For the case of the non-local $\cnot$, we have analized in full details the
theoretical procedures and the experimental requirements needed to implement the gate in our CQED system. Our analysis has shown that some practical problems have to be taken in consideration in the proposed experimental scenario. 

In particular, some difficulties arise connected with the low efficiency of the currently available sources of single-photon states operating in microwaves. On the other hand, while a local $\cnot$ $cavity\rightarrow{a}tom$ can be efficiently realized via a controlled interaction of the atom with an external field, the practical implementation of a $\cnot$ $atom\rightarrow{c}avity$ is basically an inefficient operation. This low efficiency is due to the fact that the realization of this local gate passes through an atomic transition that is forbidden by the electric-dipole-transitions selection rules. To circumvent this problem, we have proposed to set an externally driven two-photon transition between two suitably chosen states of the dressed-atom eigenspectrum. With this solution, we have found a significative improvement of the fidelity of the gate. We want to stress that the discussed difficulties are just related to the current {\it state of the art}: once these realizative prolems will be solved, our experimental proposal will certainly acquire practical reliability.

The versatility of the proposed set-up has been shown describing how to modify the protocol for a non-local $\cnot$ to get a non-local $\pi-{\gate{CQPG}}$. The realization of this gate is based on atom-external field interactions of the kind used to implement a $\cnot$ $cavity\rightarrow{a}tom$ and on resonant atom-cavity mode interactions. Because of the intrinsic high reliability of these operations, we have found that this non-local gate could be implemented in an efficient way.  

Despite the encountered difficulties to manage, the proposed set-up gives some insights in the fundamental research of possible architectures for a quantum computer able to manage large computational registers.


\section*{Acknowledgments}
We would like to thank T. Calarco, Byoung S. Ham, D. Vitali, A. Carollo and H. Jeong for
helpful discussions. This work was
supported by ESF under "Quantum Information Theory and Quantum Computing"
program, by the EU under grants
IST - EQUIP, "Entanglement in Quantum Information Processing and
Communication" and by the UK Engineering and Physical Science Research Council through GR/R33304.



\end{document}